# Dynamical Humidity Sensing Characterization of LiNbO$_3$ Film and Porous Si based Structures

S. BRAVINA[1], N. MOROZOVSKY[1], E. DOGHECHE[2], D. REMIENS[2]
and R. BOUKHERROUB[3]

[1]*Institute of Physics of NAS of Ukraine, 46 Prospect Nauki, 03028 Kiev, Ukraine*

[2]*IEMN – DOAE – MIMM Team, CNRS - UMR 8520, Bat. P3, Cité Scientifique USTL,*

*59652 Villeneuve dAscq, France*

[3]*Interdisciplinary Research Institute, IEMN-IRI, Avenue Poincaré - BP 69*

*59652 Villeneuve d'Ascq, France*

The results of the investigation of impact of pulse humidity changes on the parameters of dynamic current-voltage loops and charge-voltage loops, and also dynamic transient currents for Pt/LiNbO$_3$-film/Pt:Ti/SiO$_2$/Si and porous Si/Si structures are presented. The peculiarities of current-voltage loops and transient currents curves are characteristic for poling processes in the space charge region similar to that observed in the case of ionic semiconductors.

The efficiency of using the methods of dynamic electrophysical characterization for studying characteristics of various under fast humidity changes was demonstrated. The results are considered from the standpoint of applications in high-speed humidity sensing.

# 1. INTRODUCTION

The problem of environment parameters monitoring demands from the modern sensorics many efforts directed to development of sensitive and high-speed humidity sensors integrated in the modern silicon (Si) base.

Electrophysical and thermal characteristics of systems based on porous materials with open pores are rather sensitive to the nature of molecules penetrated into the pores [1]. That is the base of using porous materials in sensors of environmental control, in particular for humidity sensors. In this way many humidity sensitive media with open pores were examined. As promising materials are selected the following: polymers as the cheapest, porous ceramics as more flexible, and zeolite-like systems and also mezo-porous phases as thermo-stable.

Recently we have investigated the humidity-electric activity in some of porous ceramics [2] and zeolite-like systems and mezo-porous phases [3] and also porous silicon [4, 5].

Porous silicon (PSi) based structures are considered as promising for modern micro- and nano- electronics and sensorics due to its photo- and electro-luminescence properties [6] as well as due to its thermal decoupling properties [7] and humidity sensitivity [8].

The structures of $LiNbO_3$-film on Si-substrate (LN-film/Si) are considered as promising for replacement $LiNbO_3$ thin plate single crystalline elements of functional optoelectronics, in particular for surface acoustic wave (SAW) devices, non-linear optics (wave-guides and frequency doublers) and telecommunication (optical switchers and filters) applications [9, 10]. The structures of LN-film/Si are also attractive for integrated ferroelectric random access memories (FeRAM) [11].

Taking into account the well-known humidity sensing of PSi in static mode [8] and a stable tendency of integrating functional elements including ferroelectrics based structures into Si-based microelectronics that operate not only in static but also in dynamic mode, it is reasonable to perform the comparative dynamic characterization for LN-film/Si and PSi/Si structures as surface-active systems.

In this paper we present the results of comparative investigation of dynamic current-voltage and charge-voltage characteristics and also transient currents under the pulse humidity impact for LN-film/Si and PSi/Si structures.

## 2. EXPERIMENT

### 2.1. Samples

In Pt/LiNbO$_3$ film/Pt:Ti/SiO$_2$/Si-substrate (LN/Si) structures LN films were prepared by radio-frequency magnetron sputtering technique on Pt-covered Si-substrate. The procedure of deposition was similar to that of LN/Al$_2$O$_3$ structures described previously [10].

The sputtering target was a cold pressed of 75 mm of diameter pellet of commercial LiNbO$_3$ powder. The optimal growth temperature was selected 490-500 ºC for obtaining the films of single-phase composition with minimal Li-deficient. The thickness of obtained LN-film was 0,4 – 1,1 μm depending on corresponding deposition time 4 – 11 h.

The bottom Pt/Ti-bilayer was deposited onto the oxidized (350 nm of SiO$_2$) (100) n-type Si 350 μm substrate. The top Pt- electrodes with 0,15 mm$^2$ of area were deposited by sputtering procedure, which was followed by a lift-off.

The samples of PSi were prepared on 500 μm of thickness Si(111) B-doped p-type wafers with resistivity ρ = 6-10 Ω·cm. Anodic electrochemical etching was performed in electrolyte HF/EtOH=1/1 with current density of anodization 5-20 mA/cm$^2$ during 5-8 min. Then the samples were aged in the same electrolyte for 30 min. As the electrodes were used In clamped ones and Ag paste spot with area of 0,3-1 mm$^2$.

The samples of Ag-containing crystal proustite, $Ag_3AsS_3$, with mixed ion-electronic electrical conduction, that can be considered as a model material for examining phenomena connected with complex type of conductivity, were prepared from single crystal polished Z-cut plates of 0,5-0,8 mm of thickness, 3-10 mm$^2$ of area with vacuum evaporated Au electrodes.

## 2.2. Measurements

We investigated the variations of parameters of dynamic current-voltage (I-V-) and charge-voltage (Q-V-) loops and also dynamic transient currents (I-t-curves) induced by 1-3 sec of duration pulse relative humidity $H_r$ changes.

Under measuring I-V-loops and I-t-curves as a load the reference resistor was used and during the measurements of Q-V-loops that was the reference capacitor, as was done under investigations of corresponding characteristics of ferroelectric capacitors [5, 15].

The corresponding measurements were performed in the multi-cycle regime under applied a.c. triangular drive voltage (for I-V- and Q-V-loops) and meander voltage (for I-t-curves) in the frequency range 1 Hz - 1 kHz and applied voltage range of 10 mV - 10 V.

The change of sample parameters under applied voltage and/or humidity variation can be characterized by change of parameters of the equivalent series-parallel R-C-circuit with pronounced R(V) and C(V) dependences.

The main peculiarities of I-V-loops can be considered using simplified parallel R-C-circuit neglecting the effect of series resistor and capacitor. Since for this R-C-circuit $I(V) = d(CV)/dt + V/R$, under $V = V_0[1 \pm (bt - 1)]$ with $b$ = const and $C$ = const the current value is $I(V) = I(V_0, t) = V_0\{\pm bC + (1/R)[1 \pm (bt - 1)]\}$. So any deviation of I(V) from linearity is connected with existence of some R(V) and C(V) dependences.

The temporal changes of the investigated characteristics of the examined structures during and just after wet air pulse and under restoration of the initial state were also registered.

## 3. RESULTS AND COMMENTS

Figure 1 and Figure 2 present I-V- and Q-V- loops and also I-t-curves before and just after humidity pulse impact for LN-film/Si-substrate and PSi/Si structures respectively.

The shape of low-voltage parts of I-V-loops (Figures 1, 2, left) at low $H_r$ value and low-$H_r$ I-t- curves (Figures 1, 2, right) corresponds to equivalent linear series-parallel R-C-circuit.

Transformation of I-V-loops and I-t-curves under action of wet air pulse corresponds to decrease of R-value and increase of C-value under occurrence of apparent R- and C- voltage non-linearity. This state with high C- and low R- values remains during 5-20 sec depending on duration of $H_r$-pulse. Returning to the initial state is accompanied by decrease of degree of R- and C- voltage non-linearity to the initial value. At infra-low frequencies the I-V-loops show high degree of the voltage R- and C- non-linearity for both LN-film/Si and PSi/Si structures. This non-linearity is decreasing under frequency increase.

For LN-film/Si increasing $H_r$ leads to pronounce broadening of hysteresis regions on the positive and negative branches of I-V-loops (Fig. 1, left, top). This transformation is followed by appearance of the range of increasing relaxation on I-t-curves (Fig. 1, right, top). The initial shape of I-V-loops and I-t-curves restores under subsequent drying. Under short-term high humidity impact the transformation of I-V-loops to the view characteristic of threshold switching to high conductive state was observed (Fig. 1, right, bottom). Initial low conductive state is achieved after drive voltage getting off. So reversible breakdown is possible. Long-term high humidity impact leads to irreversible breakdown of the film.

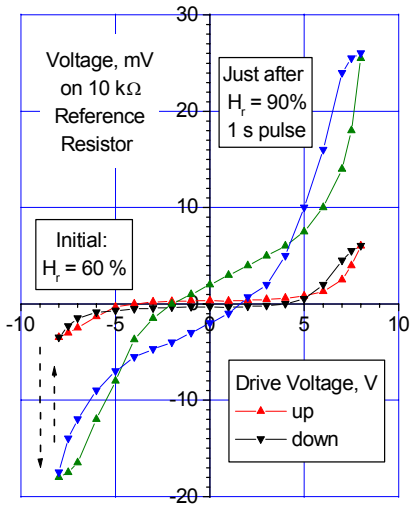
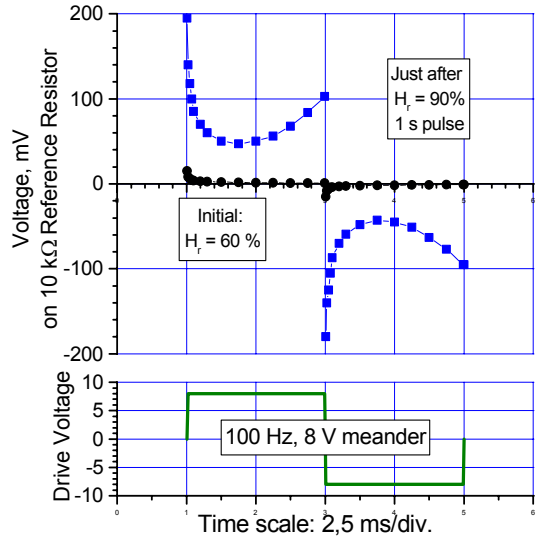
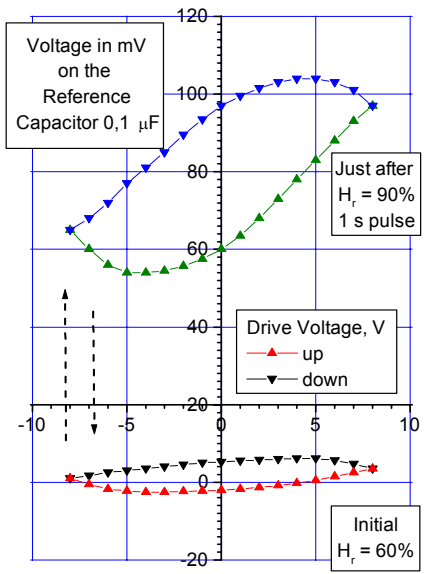
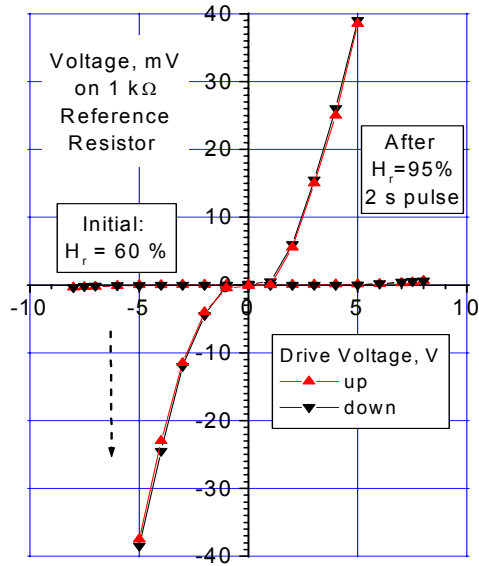

**Figure 1.** Current-voltage loops (left, top), charge-voltage loops (left, bottom) (10 Hz drive voltage) and transient currents (100 Hz drive voltage) (right, top) under low humidity impact and also current-voltage loops under high humidity impact (right, bottom) for Pt/LiNbO$_3$/Pt:Ti/SiO$_2$/Si structure before and after the pulse of relative humidity change.

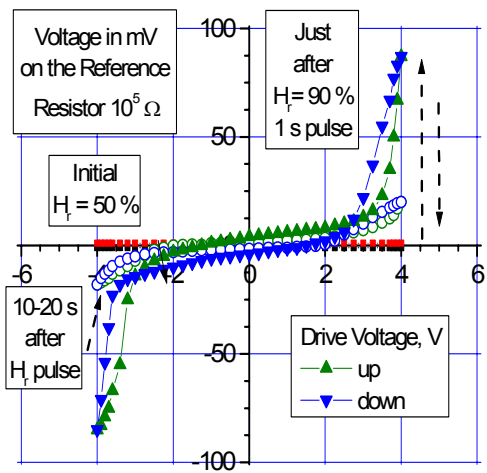
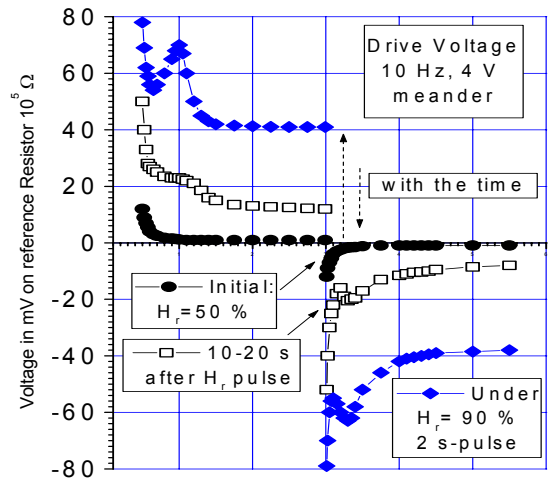
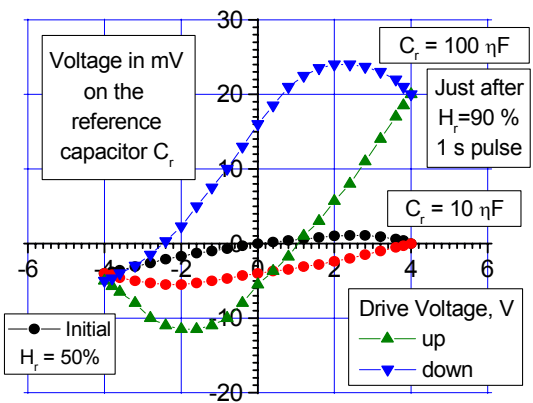

**Figure 2.** Current-voltage loops (left, top), charge-voltage loops (left, bottom) (10 Hz drive voltage) and transient currents (10 Hz, 4 V meander drive voltage) (right, top) for Ag-PSi-Ag structure before, under and after the pulse of relative humidity change.

For PSi/Si under increasing $H_r$ is observed appearance and broadening of hysteresis regions on the positive and negative branches of I-V-loops (Fig. 2, left, top) with corresponding appearance of characteristic "hump" on I-t-curves (Fig. 2, right, top). The height of this "hump" is maximal at high $H_r$ and decreases under drying PSi sample in course of returning to the initial state. The observed changes of Q-V- loops parameters under $H_r$ impact are in agreement with the same of I-V- loops and these changes are similar for LN-film/Si and PSi/Si structures (Figures 1 and 2 left, bottom).

The increase of $H_r$ leads to the increase of the loop vertical size and to the vertical shift of the loop. The first one of the observed peculiarities corresponds to the increase of the value of transferred electrical charge; the second one corresponds to predominance of the charge of definite sign in the charge transfer.

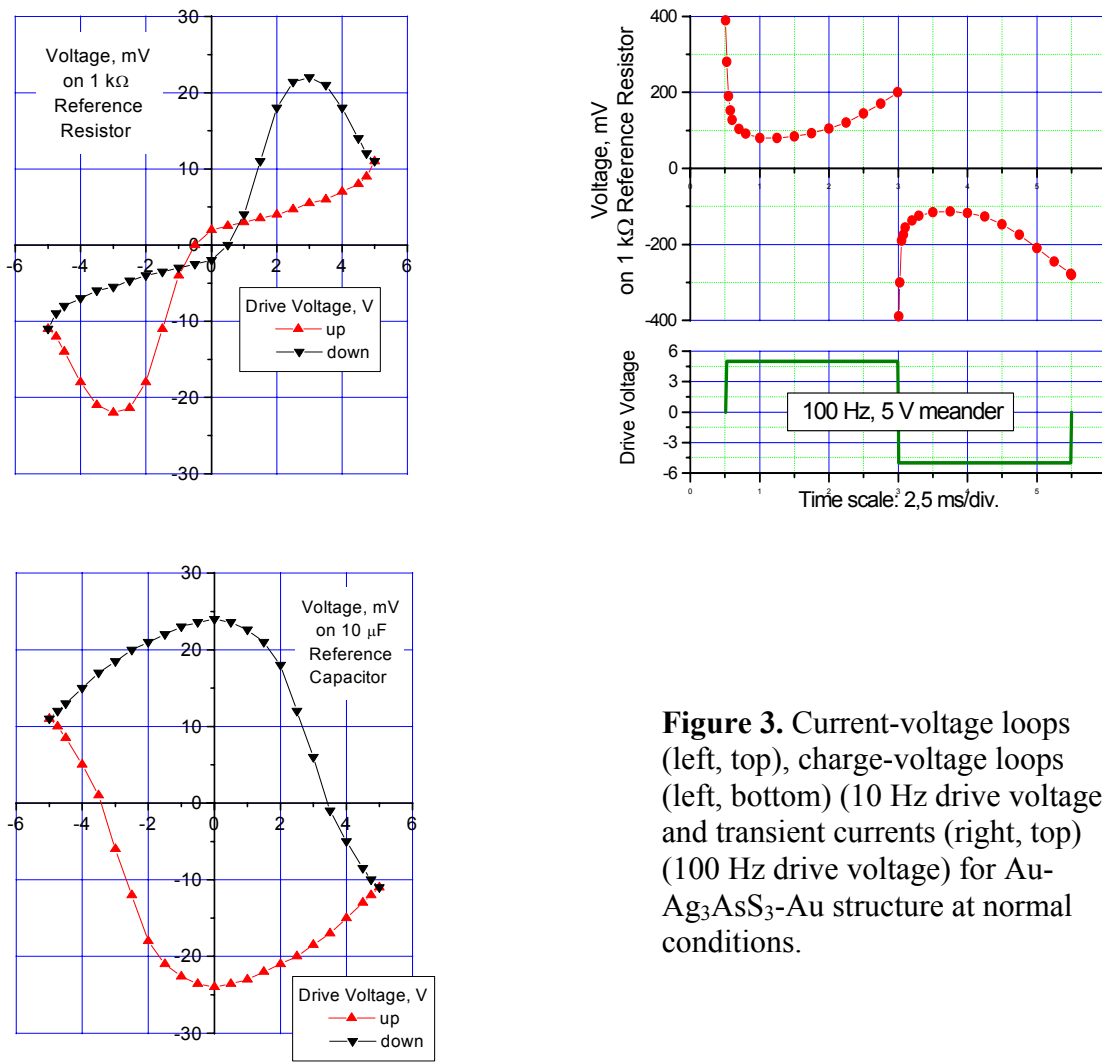

**Figure 3.** Current-voltage loops (left, top), charge-voltage loops (left, bottom) (10 Hz drive voltage) and transient currents (right, top) (100 Hz drive voltage) for Au-$Ag_3AsS_3$-Au structure at normal conditions.

Figure 3 presents I-V- and Q-V- loops and also I-t-curves for semiconductor-ionic $Ag_3AsS_3$. Specific view of these characteristics is connected with mixed $Ag^+$-$e^-$ charge transport inherent to $Ag_3AsS_3$ crystals, for which the formation of ionic space charge in undersurface region is characteristic. The shape of these I-V-loops for $Ag_3AsS_3$ and direction of path tracing are similar to those observed for humidity influenced LN-film/Si and PSi/Si structures. The shape of I-t-curves reflects similar features of time behaviour for $Ag_3AsS_3$ and humidity impacted LN-film/Si structures.

## 4. DISCUSSION

The change of the shape of I-V-loops and I-t-curves under $H_r$ variation can be characterized by linear parallel R-C-circuit for its low-voltage parts. As for its high-voltage parts it can be characterized by the series-parallel circuit with pronounced R(V) and C(V) non-linearity of all the components.

The observed transformations of Q-V- and I-V- loops are in a good correspondence. In particular, the vertical shift of Q-V-loops is connected with the rectification effect as a consequence of I-V-loops asymmetry, which leads to accumulation of d.c. electric charge on the reference capacitor during every period of a.c. driving voltage change. The indicated similarity of view of observed I-V-loops for LN-film/Si and PSi/Si and I-V- loops for the model semiconductor-ionic $Ag_3AsS_3$ points to an apparent contribution of ionic processes.

The existence of increasing relaxation on I-t-curves for LN-film/Si is similar to that observed for the model semiconductor-ionic $Ag_3AsS_3$ and $Ag_3SbS_3$ [12, 13] with the relay-race mechanism [14] of charge transfer and reaction of electrolytic type $Ag^+ + e^- = Ag^0$ in near-electrode under-surface regions.

The "humps" observed on I-t-curves for PSi/Si are characteristic for both ferroelectric systems under polarization switching which follows by displacement current pulse [15] and semiconductor systems under transfer of the packet of injected charge carriers [16].

For the investigated PSi/Si systems the observed peculiarities of Q-V- and I-V- loops are connected with adsorption of $H_2O$ vapor, post - dissociation of water molecules (H-O-H $\rightarrow$ $H^+$ + $OH^-$) and ion-electronic ($H^+$ -$e^-$ or $OH^-$-) transfer under participation of Si-SiO$_x$ conglomerates in porous media of PSi and desorption of $H_2O$ as the temporal source of ionic charge carriers. The mechanism of charge transfer in the PSi is connected with the hopping transport by means of switching dangling bonds [17]. Kinetics of these processes determines the shape of corresponding characteristics and peculiarities of their time transformation.

The results obtained for the samples PSi/Si are comparable with those obtained for porous metal-oxide ceramics and zeolite-like (of Na-Y type) and silica mezo-porous systems (of MCM-41 type) [2, 3]. These systems can be considered as a relatively inertial media for reactions connected with dissociation of water H-O-H molecules unlike the case of $LiNbO_3$-films where the electrochemical processes with participation of not only $H^+$ and $OH^-$ but also $Li^+$-sublattice carriers ($Li^+$-ion and/or $Li^-$-vacancy and also $H^+_{Li}$ hopping) are expected.

The data obtained for LN-film systems prove the great influence of the processes of absorption and dissociation of polar $H_2O$ molecules due to surface activity of the structures. Moreover, it was demonstrated possibility of not only continuous Q-V- and I-V- loops transformation (under short-term low humidity impact) but also the transformation of threshold switching type (under short-term high humidity impact) up to film breakdown (under long-term high humidity impact).

The observed time scale of transformation of I-V loops (as well as I-t curves and Q-V-loops) under humidity impact in the sequence "change of the shape – stabilization in time – shape restoration" with certain degree of approximation can be considered as corresponding to characteristic times in the sequence "adsorption - dissociation and transfer – desorption".

## 5. CONCLUSION

For PSi-layer/Si as silicon-derivative surface-active material and for LN-film/Si as an example of ferroelectric surface-active system was shown the efficiency of using the methods of dynamic electrophysical characterization for studying the changes of characteristics of these materials under fast humidity changes.

The view of high humidity current-voltage and charge-voltage loops and also current-time transient curves is characteristic for poling processes in the space charge region similar to that observed in the case of typical semiconductor-ionic materials under high enough voltage application. The observed phases of transformation of investigated electrophysical characteristics reflect the time scale of processes in the sequence "adsorption-dissociation and transfer – desorption".

The transformation of I-V-loops from continuous reversible (under low humidity impacts) up to threshold switching type (under short-term high humidity impacts) in LN/Si systems gives the possibility of consideration of such type systems as promising for fast humidity sensors of various types, in particular humidity switched sensors for high-speed humidity controlled devices of monitoring systems.

Observed humidity impact should be taken into account under operation with non-evacuated film structures based on non-ferroelectric and polar-active ferroelectric-like surface-active materials.

**ACKNOWLEDGEMENTS**

The work was supported by Ministry of Education and Science of France, Conseil Regional Nord-Pas-de-Calais and the University of Valenciennes.